\documentclass[twocolumn,amsmath,amssymb]{revtex4}
\usepackage{amsfonts}
\usepackage[usenames]{color}

\newcommand{\be}{\begin{equation}}
\newcommand{\ee}{\end{equation}}
\newcommand{\bea}{\begin{eqnarray}}
\newcommand{\eea}{\end{eqnarray}}

\begin{document}
\title{Maxwell equations in curved spacetime}
\author{Jai-chan Hwang${}^{1}$, Hyerim Noh${}^{2}$}
\address{${}^{1}$Particle Theory  and Cosmology Group,
         Center for Theoretical Physics of the Universe,
         Institute for Basic Science (IBS), Daejeon, 34126, Republic of Korea
         \\
         ${}^{2}$Theoretical Astrophysics Group, Korea Astronomy and Space Science Institute, Daejeon, Republic of Korea
         }


\begin{abstract}

In curved spacetime, Maxwell's equations can be expressed in forms valid in Minkowski background, with the effect of the metric (gravity) appearing as effective polarizations and magnetizations. The electric and magnetic (EM) fields depend on the observer's frame four-vector. We derive Maxwell's equations valid in {\it general} curved spacetime using the fields defined in the normal frame, the coordinate frame, and two other non-covariant methods used in the literature. By analyzing the case in the generic frame we show that the EM fields, as well as the charge and current densities, defined in non-covariant ways do {\it not} correspond to physical ones measured by an observer. We show that modification of the homogeneous part is inevitable to any observer, and such a modification is difficult to interpret as the effective medium property. The normal frame is the relevant one to use as it gives the EM fields measured by an Eulerian observer. 

\end{abstract}

\maketitle

%
%
%
\section{Introduction}

The special theory of relativity originated from Maxwell's electromagnetism. It showed that the electric and magnetic (EM) fields, ${\bf E}$ and ${\bf B}$, are not independent concepts, and their nature depends on the observer's motion \cite{Einstein-1905}. Minkowski, by introducing a field strength tensor $F_{ab}$, has constructed the spacetime covariant form of Maxwell's equations valid in the flat spacetime of special relativity \cite{Minkowski-1907}. With the completion of the general theory of relativity, Einstein extended the spacetime covariant form of Maxwell's equations by Minkowski, now valid in curved spacetime \cite{Einstein-1916}. In a curved spacetime, besides their dependence on the motion of the observer, the EM fields also depend on the metric, the curved nature of the spacetime, through observer's four-vector.

Using a generic timelike four-vector $u_a$, $F_{ab}$ can be decomposed into $E_a$ and $B_a$ in a spacetime covariant manner \cite{Moller-1952, Lichnerowicz-1967, Ellis-1973}; the charge and current densities, $\varrho$ and $j_a$, are similarly decomposed from the four-current $J_a$ based on the {\it same} four-vector. The four-vector $u_a$ depends on the observer. Often used ones are the fluid four-vector corresponding to the Lagrangian observer (comoving with the fluid), the normal four-vector $n_a$ corresponding to the Eulerian observer, the coordinate four-vector $\bar n_a$ corresponding to an observer attached with spatial coordinates, the coordinate observer; in the following, the normal four-vector, the normal frame, and the normal (or Eulerian) observer are used with the same meaning indicating $n_a$.

In this work, we will consider the normal observer, the coordinate observer, and the all encompassing generic observer. As further two cases, we consider non-covariant definitions of the EM fields, which are historically interesting, partly coincide with flat spacetime results, and are often used in the literature related to the gravitational wave detection and the gravity-medium analogy. However, these two non-covariant definitions fail to identify the corresponding observer's four-vectors, and thus fail to introduce the external charge and current densities properly \cite{Hwang-Noh-2023-EM-definition}. We will show this failure by analysing the generic frame; i.e., the non-covariant definitions do not have corresponding frame four-vectors. Without the observer's four-vector identified, the actual existence of such a set of EM fields as measurable quantities by any observer is in doubt. This also implies that the homogeneous part of Maxwell's equations is inevitably modified by gravity to any observer. Thus, the gravity-medium analogy is not valid.

In \cite{Hwang-Noh-2023-EM-definition}, we derived Maxwell's equations in the above four different definitions of the EM fields, assuming {\it linear} order perturbations in the Minkowski background metric, thus assuming weak gravity. Maxwell's equations were presented in the Minkowski background, with the effect of weak gravity appearing as the effective polarizations and magnetizations, ${\bf P}$s and ${\bf M}$s, often appearing in the homogeneous and/or inhomogeneous parts of Maxwell's equations. We showed why the non-covariant definitions of the EM fields are not only arbitrary but also wrong by missing the corresponding charge and current densities and missing the observer who can measure such fields. That is, the variables introduced in that manner are {\it not} EM fields in any sense.

Here we extend the case to the {\it general} spacetime metric. Equations will be presented using the EM fields and current density associated with the Euclidean three-space metric $\delta_{ij}$ mixed with the Arnowitt-Deser-Misner (ADM) metric variables. The ADM metric variables can be expressed in terms of the fully nonlinear and exact (FNLE) perturbation metric variables where vector and tensor metric perturbation variables are associated with the Euclidean three-space metric $\delta_{ij}$ and its inverse. We additionally present the case of the generic four-vector and its weak gravity limit.

In Sec.\ \ref{sec:covariant-ADM} we present the spacetime covariant formulation and the spatially covariant ADM formulation of Maxwell's equations. Section \ref{sec:Maxwell} presents Maxwell's equations in four different definitions of the EM fields. Section \ref{sec:generic} presents the equations in the generic frame and proves the inevitable nature of the gravity-modified homogeneous part. In Secs.\ \ref{sec:misconception} and \ref{sec:medium} we clarify a couple of misconceptions in the literature and errors in the commonly used medium interpretation of gravity in electrodynamics. In Sec.\ \ref{sec:Discussion}, we discuss the result. Appendix \ref{sec:FNLE} presents the ADM metric variables in terms of variables associated with $\delta_{ij}$ and its inverse using the FNLE formulation. Appendix \ref{sec:linear} presents relations to the linear order metric perturbation which will be useful to handle electrodynamics coupled with gravitational waves and the weak gravity.

%
%
%
\section{Covariant and ADM formulations}
                                      \label{sec:covariant-ADM}

In the presence of helical coupling, the electromagnetic part of Lagrangian is
\bea
   L = - {1 \over 4} F_{ab} F^{ab}
       - {g_{\phi \gamma} \over 4} f (\phi) F_{ab} F^{*ab}
       + {1 \over c} J^a A_a,
\eea
where the field strength tensor is $F_{ab} \equiv \nabla_a A_b - \nabla_b A_a$ with its dual $F^*_{ab} \equiv {1 \over 2} \eta_{abcd} F^{cd}$; $\phi$ is a scalar field with $f = \phi$ for the conventional axion coupling \cite{Sikivie-1983}. A system together with Einstein's gravity, scalar field, and a general fluid is studied in \cite{Hwang-Noh-2022-Axion-EM}.

Maxwell's equations in vacuum are
\bea
   F^{ab}_{\;\;\;\;;b}
       = {1 \over c} J^a - g_{\phi\gamma} f_{,b} F^{*ab}, \quad
       \eta^{abcd} F_{bc,d} = 0.
   \label{Maxwell-tensor-eqs}
\eea
The second equation is the same as $F^{*ab}_{\;\;\;\;\;\;;b} = 0$. Using it, the first one becomes
\bea
   H^{ab}_{\;\;\;\;;b}
       = {1 \over c} J^a,
   \label{Hab-eq}
\eea
with $H_{ab} \equiv F_{ab} + g_{\phi\gamma} f F^*_{ab}$. Using $H^{ab} = - {1 \over 2} \eta^{abcd} H^*_{cd}$ this can be written as
\bea
   - {1 \over 2} \eta^{abcd} H^*_{bc,d} = {1 \over c} J^a.
   \label{Hab*-eq}
\eea

Using a generic time-like four-vector $u_a$, with $u_a u^a \equiv -1$, we {\it define} the EM fields associated with the four-vector \cite{Moller-1952, Lichnerowicz-1967, Ellis-1973}
\bea
   & & F_{ab} = u_a E_b - u_b E_a
       - \eta_{abcd} u^c B^d,
   \nonumber \\
   & & F^*_{ab} = u_a B_b - u_b B_a
       + \eta_{abcd} u^c E^d.
   \label{Fab-cov}
\eea
We have $E_a \equiv F_{ab} u^b$ and $B_a \equiv F_{ab}^* u^b$ with $E_a u^a \equiv 0 \equiv B_a u^a$. Thus, we have
\bea
   & & H_{ab} = u_a D_b - u_b D_a
       - \eta_{abcd} u^c H^d,
   \nonumber \\
   & & H^*_{ab} = u_a H_b - u_b H_a
       + \eta_{abcd} u^c D^d.
   \label{Hab-cov}
\eea
with
\bea
   & & D_a \equiv E_a + P^{\rm A}_a, \quad
       P^{\rm A}_a \equiv g_{\phi\gamma} f B_a,
   \nonumber \\
   & &
       H_a \equiv B_a - M^{\rm A}_a, \quad
       M^{\rm A}_a \equiv g_{\phi\gamma} f E_a.
\eea
The effect of axion can be regarded as the effective polarization and magnetization. The current four-vector is decomposed using the {\it same} four-vector as
\bea
   J^a \equiv \varrho c u^a
       + j^a, \quad
       j_a u^a \equiv 0,
   \label{J^a}
\eea
where $\varrho$ and $j_a$ are charge and current densities, respectively, measured by an observer with the same four-velocity $u_a$.

Using the decompositions, Eq.\ (\ref{Hab-eq}) and the second one in Eq.\ (\ref{Maxwell-tensor-eqs}), respectively, give
\bea
   & & \hskip -.8cm
       {1 \over \sqrt{-g}} \left[ \sqrt{-g}
       ( u^a D^b - u^b D^a ) \right]_{,b}
       - \eta^{abcd} ( u_c H_d )_{,b}
   \nonumber \\
   & & \qquad \hskip -.8cm
       = \varrho u^a + {1 \over c} j^a,
   \label{Maxwell-DH} \\
   & & \hskip -.8cm
       {1 \over \sqrt{-g}} \left[ \sqrt{-g}
       ( u^a B^b - u^b B^a ) \right]_{,b}
       + \eta^{abcd} ( u_c E_d )_{,b} = 0,
   \label{Maxwell-EB}
\eea
where $g \equiv {\rm det}(g_{ab})$. As we have $\eta^{abcd} = \epsilon^{abcd}/\sqrt{-g}$ with $\epsilon^{abcd}$ an anti-symmetric symbol, Maxwell's equations can be written in forms with the effect of gravity (curved spacetime metric) behaving as the effective polarizations and magnetizations appearing in the homogeneous and inhomogeneous parts for a generic four-vector $u_a$. In the following, we will show that modification in the homogeneous part due to gravity is inevitable to any observer even to the linear order perturbation. From now on, for clarity, we often indicate the spacetime covariant quantities using an overtilde.

We often use spatially covariant formulation of the curved spacetime. The ADM metric and its inverse are \cite{ADM}
\bea
   & & g_{00} \equiv - N^2 + N^i N_i, \quad
       g_{0i} \equiv N_i, \quad
       g_{ij} \equiv \overline h_{ij},
   \nonumber \\
   & & g{}^{00} = - {1 \over N^2}, \quad
       g{}^{0i} = {N^i \over N^2}, \quad
       g{}^{ij} = \overline h{}^{ij}
       - {N^i N^j \over N^2},
   \label{ADM-metric}
\eea
where $\overline h{}^{ij}$ is an inverse of the three-space intrinsic metric $\overline h_{ij}$, thus $\overline h{}^{ik} \overline h_{jk} \equiv \delta^i_j$, and the index of $N_i$ is raised and lowered by $\overline h_{ij}$ and its inverse; $i, j \dots$ are spatial indices. We have
\bea
   & & \eta_{0ijk} = - N \overline \eta_{ijk}
       = - N \sqrt{\overline h} \eta_{ijk}
       = - \sqrt{- g} \eta_{ijk},
   \nonumber \\
   & & \eta{}^{0ijk}
       = {1 \over N} \overline \eta{}^{ijk}
       = {1 \over N \sqrt{\overline h}} \eta^{ijk}
       = {1 \over \sqrt{- g}} \eta^{ijk},
   \label{eta-definition}
\eea
where $\overline h \equiv {\rm det}(\overline h_{ij})$, and indices of $\overline \eta_{ijk}$ and $\eta_{ijk}$ are raised and lowered using $\overline h_{ij}$ and $\delta_{ij}$, respectively, and their respective inverses.

The normal four-vector is introduced as
\bea
   n_i \equiv 0, \quad
       n_0 = - N, \quad
       n{}^i = - {N^i \over N}, \quad
       n{}^0 = {1 \over N}.
   \label{n_a-ADM}
\eea
For the EM fields and current density in the normal frame, we set
\bea
   \widetilde B_i \equiv \overline B_i, \quad
       \widetilde B_0 = \overline B_i N^i, \quad
       \widetilde B^i = \overline B{}^i, \quad
       \widetilde B^0 = 0,
\eea
and similarly for $\overline E_a$, $\overline D_a$ $\overline H_a$, and $\overline j_a$ with $\widetilde E_i \equiv \overline E_i$ etc.; indices of $\overline E_i$, etc., are raised and lowered by $\overline h_{ij}$ and its inverse. Based on the normal frame, from Eq.\ (\ref{Fab-cov}) we have
\bea
   & & F_{0i} = - N \overline E_i
       - \overline \eta_{ijk} N^j \overline B{}^k, \quad
       F_{ij} = \overline \eta_{ijk} \overline B{}^k,
   \nonumber \\
   & & F{}^*_{0i} = - N \overline B_i
       + \overline \eta_{ijk} N^j \overline E{}^k, \quad
       F{}^*_{ij} = - \overline \eta_{ijk}
       \overline E{}^k,
   \label{Fab-ADM}
\eea
and similarly for $H_{ab}$ and $H^*_{ab}$ with ($\overline D_i$, $\overline H_i$) replacing ($\overline E_i$, $\overline B_i$).

Maxwell's equations in the ADM formulation are presented in textbooks of numerical relativity \cite{Baumgarte-Shapiro-2010, Gourgoulhon-2012, Shibata-2015}; see Eqs.\ (65)-(73) in \cite{Hwang-Noh-2022-Axion-EM}. Equations can be arranged, including the axion coupling, as
\bea
   & & \hskip -.8cm
       {1 \over \sqrt{\overline h}} \left( \sqrt{\overline h}
       \overline D^i \right)_{,i}
       = \varrho,
   \label{Maxwell-n-1} \\
   & & \hskip -.8cm
       {1 \over \sqrt{\overline h}} \left( \sqrt{\overline h}
       \overline D^i \right)_{,0}
       - \overline \eta^{ijk} \overline \nabla_j
       ( N \overline H_k - \overline \eta_{k\ell m} N^\ell
       \overline D{}^m )
   \nonumber \\
   & & \qquad \hskip -.8cm
       = \varrho N^i
       - {1 \over c} N \overline j{}^i,
   \label{Maxwell-n-2} \\
   & & \hskip -.8cm
       {1 \over \sqrt{\overline h}} \left( \sqrt{\overline h}
       \overline B^i \right)_{,i} = 0,
   \label{Maxwell-n-3} \\
   & & \hskip -.8cm
       {1 \over \sqrt{\overline h}} \left( \sqrt{\overline h}
       \overline B^i \right)_{,0}
       + \overline \eta^{ijk} \overline \nabla_j
       ( N \overline E_k + \overline \eta_{k\ell m} N^\ell
       \overline B{}^m ) = 0.
   \label{Maxwell-n-4}
\eea
Using the vector notation, we have
\bea
   & & \hskip -.8cm
       \overline \nabla \cdot \overline {\bf D}
       = \varrho,
   \\
   & & \hskip -.8cm
       {1 \over \sqrt{\overline h}} \big( \sqrt{\overline h}
       \overline {\bf D} \big)_{,0}
       - \overline \nabla \times \left( N \overline {\bf H}
       - {\bf N} \times \overline {\bf D}
       \right)
       = {\bf N} \varrho
       - {1 \over c} N \overline {\bf j},
   \\
   & & \hskip -.8cm
       \overline \nabla \cdot \overline {\bf B}
       = 0,
   \\
   & & \hskip -.8cm
       {1 \over \sqrt{\overline h}} \big( \sqrt{\overline h}
       \overline {\bf B} \big)_{,0}
       + \overline \nabla \times \left( N \overline {\bf E}
       + {\bf N} \times \overline {\bf B}
       \right)
       = 0,
\eea
where the dot and cross products are associated with the intrinsic metric $\overline h_{ij}$, and $\overline \nabla$ is a covariant derivative associated with $\overline h_{ij}$. 

\section{Maxwell's equations}
                                        \label{sec:Maxwell}

In this section, we present exact forms of Maxwell's equations in the general curved spacetime using the EM field vectors associated with $\delta_{ij}$ as the metric. We present four different forms based on four different ways of defining the EM fields. Two covariant methods are based on the normal frame and the coordinate frame as the observer's four-vectors. The other two non-covariant methods are based on the coincidences with the special relativistic equations noticed by Einstein in his way to guess the spacetime covariant form of Maxwell's equation valid in general relativity \cite{Einstein-1916}; these non-covariant methods are also popularly used in the literature.

The two non-covariant definitions are not based on the observer's four-vector: the EM fields are directly read as components of special relativistic forms of $F_{ab}$ and $F^*_{ab}$, both with covariant indices. It turns out that, in this way, one cannot identify the frame four-vectors allowing such definitions of the EM fields, see Sec.\ \ref{sec:generic}. The situations are troublesome because without the observer's four-vectors specified, one cannot introduce the charge and current densities associated with the respective EM fields. Without the associated four-vector identified, the EM variables introduced in non-covariant manner are {\it not} the EM fields that can be measured by any observer.

\subsection{Normal observer}

The normal four-vector $n_a$ is normal to the hypersurface and is the four-velocity of an observer instantaneously at rest in the chosen time slice. It can be interpreted as an Eulerian observer as its motion follows the hypersurface independently of the coordinates chosen \cite{Smarr-York-1978, Wilson-Mathews-2003, Gourgoulhon-2012}.

We introduce
\bea
   \overline B_i \equiv B_i, \quad
       \overline B^i = \overline h{}^{ij} \overline B_j
       = \overline h{}^{ij} B_j,
\eea
or directly,
\bea
   \widetilde B_i \equiv B_i, \quad
       \widetilde B_0 = B_i N^i, \quad
       \widetilde B^i = \overline h^{ij} B_j, \quad
       \widetilde B^0 = 0,
\eea
where the index of $B_i$ is raised and lowered using $\delta_{ij}$ and its inverse; similarly we introduce $E_i$, $D_i$, $H_i$, and $j_i$. Equations (\ref{Fab-cov}) and (\ref{J^a}) give
\bea
   & & \hskip -.5cm
       F_{0i} = - N E_i
       - \eta_{ijk} \sqrt{\overline h} N^j
       \overline h{}^{k\ell} B_\ell, \quad
       F_{ij} = \eta_{ijk} \sqrt{\overline h}
       \overline h{}^{k\ell} B_\ell,
   \nonumber \\
   & & \hskip -.5cm
       J^0 = \varrho c {1 \over N},
       \quad
       J^i = - \varrho c {N^i \over N}
       + \overline h{}^{ij} j_j,
   \label{Fab-normal}
\eea
and similarly for $H_{ab}$. Notice that we are using mixed notation associated with $\delta_{ij}$ and $\overline h_{ij}$ and their respective inverses: the indices of $N^i$ and $\overline h{}^{ij}$ are associated with the intrinsic metric $\overline h_{ij}$, whereas indices of $\eta_{ijk}$, $E_i$, etc., are raised and lowered using $\delta_{ij}$ and its inverse. In Appendix \ref{sec:FNLE} we will express the ADM metric variables in terms of metric perturbation variables associated with $\delta_{ij}$ and its inverse. In this way, all variables in the following can be expressed entirely using variables associated with $\delta_{ij}$.

Using Eq.\ (\ref{Fab-normal}), Eq.\ (\ref{Maxwell-tensor-eqs}) gives
\bea
   & & ( \sqrt{\overline h} \overline h{}^{ij} D_j )_{,i}
       = \sqrt{\overline h} \varrho,
   \label{Maxwell-normal-1-1} \\
   & &
       ( \sqrt{\overline h} \overline h{}^{ij} D_j )_{,0}
       - \eta^{ijk} \nabla_j
       ( N H_k
       - \eta_{k\ell m} \sqrt{\overline h} N^\ell
       \overline h{}^{mn} D_n )
   \nonumber \\
   & & \qquad
       = \sqrt{\overline h} N^i \varrho
       - {1 \over c} N \sqrt{\overline h} \overline h{}^{ij} j_j,
   \label{Maxwell-normal-1-2} \\
   & & ( \sqrt{\overline h} \overline h{}^{ij} B_j )_{,i} = 0,
   \label{Maxwell-normal-1-3} \\
   & & ( \sqrt{\overline h} \overline h{}^{ij} B_j )_{,0}
       + \eta^{ijk} \nabla_j ( N E_k
       + \eta_{k\ell m} \sqrt{\overline h} N^\ell
       \overline h{}^{mn} B_n )
   \nonumber \\
   & & \qquad
       = 0,
   \label{Maxwell-normal-1-4}
\eea
where $\nabla_i$ is an ordinary derivative associated with $\delta_{ij}$. Useful relations needed for the derivation can be found in \cite{Hwang-Noh-2022-Axion-EM}. In terms of the effective polarization and magnetization vectors, these can be arranged as
\bea
   & & ( E^i + P_{\rm E}^i )_{,i}
       = \sqrt{\overline h} \varrho,
   \label{Maxwell-normal-1} \\
   & &
       ( E^i + P_{\rm E}^i )_{,0}
       - \eta^{ijk} \nabla_j
       ( B_k - M^{\rm E}_k )
   \nonumber \\
   & & \qquad
       = \sqrt{\overline h} N^i \varrho
       - {1 \over c} N \sqrt{\overline h} \overline h{}^{ij} j_j,
   \label{Maxwell-normal-2} \\
   & & ( B^i + P_{\rm B}^i )_{,i} = 0,
   \label{Maxwell-normal-3} \\
   & & ( B^i + P_{\rm B}^i )_{,0}
       + \eta^{ijk} \nabla_j ( E_k - M^{\rm B}_k ) = 0,
   \label{Maxwell-normal-4}
\eea
where the effective ${\bf P}$s and ${\bf M}$s caused by the metric are
\bea
   & & P_{\rm E}^i
       \equiv \sqrt{\overline h} \overline h{}^{ij} D_j - E^i,
   \nonumber \\
   & & M^{\rm E}_i
       \equiv B_i - N H_i
       + \eta_{ijk} \sqrt{\overline h} N^j \overline h{}^{k\ell} D_\ell,
   \nonumber \\
   & & P_{\rm B}^i
       \equiv \sqrt{\overline h} \overline h{}^{ij} B_j - B^i,
   \nonumber \\
   & & M^{\rm B}_i
       \equiv (1 - N) E_i
       - \eta_{ijk} \sqrt{\overline h} N^j \overline h{}^{k\ell} B_\ell.
   \label{PM-normal}
\eea
To the normal frame observer, the gravitational field appears as effective ${\bf P}$s and ${\bf M}$s in {\it both} the homogeneous and inhomogeneous parts of Maxwell's equations.

\subsection{Coordinate observer}
                                        \label{sec:coordinate}

In the literature, we often find a frame choice with $u^i \equiv 0$ \cite{Wilson-Mathews-2003, Baumgarte-Shapiro-2010, deFelice-1971}. This corresponds to the coordinate frame where the observer is attached with the coordinate, see below Eq.\ (\ref{Lorentz}). In the coordinate frame, we have
\bea
   & & \bar n_i = {N_i \over \sqrt{N^2 - N^k N_k}}, \quad
       \bar n_0 = - \sqrt{N^2 - N^k N_k},
   \nonumber \\
   & &
       \bar n^i \equiv 0, \quad
       \bar n^0 = {1 \over \sqrt{N^2 - N^k N_k}},
\eea
and the observer is at rest in the spatial coordinate, thus is a coordinate observer.

For the EM fields and current density in the coordinate frame, we set
\bea
   & & \widetilde B_i \equiv \bar B_i, \quad
       \widetilde B_0 = 0,
   \nonumber \\
   & &
       \widetilde B^i = \left( \overline h{}^{ij}
       - {N^i N^j \over N^2} \right) \bar B_j, \quad
       \widetilde B^0 = {N^i \over N^2} \bar B_i,
\eea
and similarly for $\bar E_a$, $\bar D_a$, $\bar H_a$, and $\bar j_a$ with $\widetilde E_i \equiv \bar E_i$ etc.; indices of $\bar B_i$, etc., are raised and lowered using $\delta_{ij}$ and its inverse. Equations (\ref{Fab-cov}) and (\ref{J^a}) give
\begin{widetext}
\bea
   & & F_{0i} = - \sqrt{N^2 - N^k N_k} \bar E_i, \quad
       F_{ij} = {1 \over \sqrt{N^2 - N^m N_m}}
       \left[ N_i \bar E_j - N_j \bar E_i
       + \eta_{ijk} N \sqrt{\overline h}
       \left( \overline h{}^{k\ell}
       - {N^k N^\ell \over N^2} \right) \bar B_\ell \right],
   \nonumber \\
   & & J^0 = {1 \over \sqrt{N^2 - N^k N_k}}
       \bar \varrho c
       + {N^i \over N^2} \bar j_i, \quad
       J^i = \left( \overline h{}^{ij}
       - {N^i N^j \over N^2} \right) \bar j_j,
\eea
and similarly for $H_{ab}$. Maxwell's equations follow from Eq.\ (\ref{Maxwell-tensor-eqs}) as
\bea
   & & \bigg\{ {N \sqrt{\overline h}
       \over \sqrt{N^2 - N^\ell N_\ell}}
       \left[ \left( \overline h{}^{ij}
       - {N^i N^j \over N^2} \right) \bar D_j
       - \eta^{ijk}{1 \over N \sqrt{\overline h}}
       N_j \bar H_k \right] \bigg\}_{,i}
       = \sqrt{\overline h}
       \bigg( {N \over \sqrt{N^2 - N^k N_k}}
       \bar \varrho
       + {1 \over c} {N^i \over N} \bar j_i \bigg),
   \label{Maxwell-coordinate-1} \\
   & & \bigg\{ {N \sqrt{\overline h}
       \over \sqrt{N^2 - N^\ell N_\ell}}
       \left[ \left( \overline h{}^{ij}
       - {N^i N^j \over N^2} \right) \bar D_j
       - \eta^{ijk} {1 \over N \sqrt{\overline h}}
       N_j \bar H_k \right] \bigg\}_{,0}
       - \eta^{ijk} \nabla_j
       \left( \sqrt{N^2 - N^\ell N_\ell} \bar H_k \right)
   \nonumber \\
   & & \qquad
       = - {1 \over c} N \sqrt{\overline h}
       \left( \overline h{}^{ij} - {N^i N^j \over N^2} \right)
       \bar j_j,
   \label{Maxwell-coordinate-2} \\
   & & \bigg\{ {N \sqrt{\overline h}
       \over \sqrt{N^2 - N^\ell N_\ell}}
       \left[ \left( \overline h{}^{ij}
       - {N^i N^j \over N^2} \right) \bar B_j
       + \eta^{ijk} {1 \over N \sqrt{\overline h}}
       N_j \bar E_k \right] \bigg\}_{,i}
       = 0,
   \label{Maxwell-coordinate-3} \\
   & & \bigg\{ {N \sqrt{\overline h}
       \over \sqrt{N^2 - N^\ell N_\ell}}
       \left[ \left( \overline h{}^{ij}
       - {N^i N^j \over N^2} \right) \bar B_j
       + \eta^{ijk} {1 \over N \sqrt{\overline h}}
       N_j \bar E_k \right] \bigg\}_{,0}
       + \eta^{ijk} \nabla_j
       \left( \sqrt{N^2 - N^\ell N_\ell} \bar E_k \right)
       = 0.
   \label{Maxwell-coordinate-4}
\eea
Similarly as in Eqs.\ (\ref{Maxwell-normal-1})-(\ref{Maxwell-normal-4}), these equations can also be expressed using effective $\bar {\bf P}$s and $\bar {\bf M}$s appearing in both the homogeneous and inhomogeneous parts of Maxwell's equations.

Although we suuggest the normal frame as the one to use because it corresponds to the frame of an Eulerian observer, as both frames are introduced in covariant manner, any frame is fine mathematically. Relations among parameters (the EM fields and charge and current densities) between the two frames can be easily read by evaluating Eqs.\ (\ref{Fab-cov}) and (\ref{J^a}) in both frames. These are
\bea
   & & E_i = {N \over \sqrt{N^2 - N^m N_m}}
       \bigg[ \left( \delta_i^j - {N_i N^j \over N^2} \right)
       \bar E_j
       - \eta_{ijk} {\sqrt{\overline h} \over N}
       N^j \overline h{}^{k\ell} \bar B_\ell \bigg],
   \nonumber \\
   & & B_i = {N \over \sqrt{N^2 - N^m N_m}} \bigg[
       \left( \delta_i^j - {N_i N^j \over N^2} \right) \bar B_j
       + \eta_{ijk} {\sqrt{\overline h} \over N}
       N^j \overline h{}^{k\ell} \bar E_\ell \bigg],
   \nonumber \\
   & & \bar E_i = {N \over \sqrt{N^2 - N^m N_m}}
       \bigg( E_i + \eta_{ijk} {\sqrt{\overline h} \over N}
       N^j \overline h{}^{k \ell} B_\ell \bigg),
   \nonumber \\
   & & \left( \delta_i^j - {N_i N^j \over N^2} \right) \bar B_j
       = {N \over \sqrt{N^2 - N^m N_m}} \bigg[
       \left( \delta_i^j - {N_i N^j \over N^2} \right) B_j
       - \eta_{ijk} {\sqrt{\overline h} \over N}
       N^j \overline h{}^{k\ell} E_\ell \bigg];
   \nonumber \\
   & & \varrho = \bar \varrho {N \over \sqrt{N^2 - N^k N_k}}
       + {1 \over c} {N^i \over N} \bar j_i, \quad
       j_i = \bar j_i
       + \bar \varrho c {N_i \over \sqrt{N^2 - N^k N_k}},
   \nonumber \\
   & & \bar \varrho = {N \over \sqrt{N^2 - N^k N_k}}
       \left( \varrho - {1 \over c} {N^i \over N} j_i \right), \quad
       \left( \delta_i^j - {N_i N^j \over N^2} \right) \bar j_j
       = j_i - \varrho c {N_i \over N},
\eea
\end{widetext}
\noindent
and similarly between ($D_i$, $H_i$) and ($\bar D_i$, $\bar H_i$). The coordinate frame coincides with the normal frame for $N_i = 0$.

\subsection{Special relativistic $F_{ab}$ or $F{}^*_{ab}$}

In this subsection we {\it ignore} the axion contribution for clarity of the argument. In his way of arriving at Eq.\ (\ref{Maxwell-tensor-eqs}) in general relativity, in \cite{Einstein-1916} Einstein noticed a couple of exact coincidences with the special relativistic Maxwell's equations based on two different definitions of the EM fields. These are
\bea
       F_{0i}
       \equiv - {\hat E}_i, \quad
       F_{ij}
       \equiv \eta_{ijk} \hat B{}^k,
   \label{Fab-hat}
\eea
and
\bea
       F{}^*_{0i}
       \equiv - {\breve B}_i, \quad
       F{}^*_{ij}
       \equiv - \eta_{ijk} \breve E{}^k,
   \label{Fab-breve}
\eea
where indices of ${\hat E}_i$, ${\hat B}_i$, ${\breve E}_i$, and ${\breve B}_i$ are raised and lowered using $\delta_{ij}$ and its inverse. Notice that these definitions of EM fields, directly associating the physical EM fields to components of tensors, are {\it not} covariant. The merit of these two definitions noticed by Einstein \cite{Einstein-1916} is that, in these ways, the second part of Eq.\ (\ref{Maxwell-tensor-eqs}) using Eq.\ (\ref{Fab-hat}) and Eq.\ (\ref{Hab*-eq}) using Eq.\ (\ref{Fab-breve}), in the absence of the external current four-vector, naturally gives the homogeneous part of Maxwell's equations in exactly special relativistic form as
\bea
   & & {\breve E}^i_{\;\;,i}
       = {1 \over c} \sqrt{-g} J^0,
   \label{Maxwell-mixed-1} \\
   & & {\breve E}^i_{\;\;,0}
       - \eta^{ijk} \nabla_j {\breve B}_k
       = - {1 \over c} \sqrt{-g} J^i,
   \label{Maxwell-mixed-2} \\
   & & {\hat B}^i_{\;\;,i} = 0,
   \label{Maxwell-mixed-3} \\
   & & {\hat B}^i_{\;\;,0}
       + \eta^{ijk} \nabla_j {\hat E}_k = 0.
   \label{Maxwell-mixed-4}
\eea
This set is a combination of parts of two sets of Maxwell's equations associated with two different definitions of the EM fields in Eqs.\ (\ref{Fab-hat}) and (\ref{Fab-breve}). The more complicated complementary parts are ignored, see below.

Notice that for the external source terms, we kept the components of the four-current without decomposing it into the charge and current densities. The two non-covariant ways of defining EM fields in Eqs.\ (\ref{Fab-hat}) and (\ref{Fab-breve}) do not guarantee the presence of the corresponding observer's four-vectors: the normalized four-vector only has three degrees of freedom, whereas in Eq.\ (\ref{Fab-hat}) or Eq.\ (\ref{Fab-breve}) six degrees of freedom are used. It turns out that even to the linear order perturbations in the metric we {\it cannot} find the four-vectors for the EM fields satisfying these conditions, see \cite{Hwang-Noh-2023-EM-definition}. Without the observer's four-vector, we cannot introduce the charge and current densities associated with the observer, which is true even for the EM fields as well. For further trouble with these non-covariant ways of identifying physical quantities as tensor components, see \cite{Crater-1994}.

Comparing Eq.\ (\ref{Fab-normal}) with Eqs.\ (\ref{Fab-hat}) and (\ref{Fab-breve}) we have relations to EM fields in the normal frame
\bea
   & & \hat E_i = N E_i
       + \eta_{ijk} \sqrt{\overline h} N^j \overline h{}^{k\ell} B_\ell
       = E_i - M_i^{\rm B},
   \nonumber \\
   & & \hat B^i = \sqrt{\overline h} \overline h{}^{ij} B_j
       = B^i + P^i_{\rm B}.
   \nonumber \\
   & & {\breve E}^i = \sqrt{\overline h} \overline h{}^{ij} E_j
       = E^i + P^i_{\rm E},
   \nonumber \\
   & & {\breve B}_i = N B_i
       - \eta_{ijk} \sqrt{\overline h} N^j \overline h{}^{k\ell} E_\ell
       = B_i - M_i^{\rm E}.
   \label{EB-relation-normal}
\eea
These can be inverted to give
\bea
   & & E_i = {1 \over N} ( \hat E_i
       - \eta_{ijk} N^j \hat B{}^k ), \quad
       B_i = {1 \over \sqrt{\overline h}} \overline h_{ij} \hat B^j,
   \nonumber \\
   & & E_i = {1 \over \sqrt{\overline h}}
       \overline h_{ij} \breve E{}^j, \quad
       B_i = {1 \over N} ( \breve B_i
       + \eta_{ijk} N^j \breve E{}^k ).
   \label{EB-relation-normal-2}
\eea

Equations (\ref{Maxwell-mixed-1})-(\ref{Maxwell-mixed-4}) look deceivingly simple: in the absence of the four-current, and if we ignore (we cannot, of course) the difference between ($\hat E_i$, $\hat B_i$) and ($\breve E_i$, $\breve B_i$), these are exactly the same as Maxwell's equations in special relativity. If we express the two sets using consistent notations, the other parts of equations become quite complicated. From
\bea
   F{}^*_{ab}
       = {1 \over 2} \eta_{abcd}
       g{}^{ce} g{}^{df} F_{ef},
\eea
and its inverse relation, we can derive the relation between the two EM fields
\bea
   & & \breve E^i = {\sqrt{\overline h} \over N}
       \overline h{}^{ij}
       \left( \hat E_j - \eta_{jk\ell} N^k \hat B{}^\ell \right)
       \equiv \hat E^i + \hat P^i_{\rm E},
   \nonumber \\
   & & \breve B_i = {N \over \sqrt{\overline h}}
       \left[ \left( 1 - {N^k N_k \over N^2} \right) \overline h_{ij}
       + {N_i N_j \over N^2} \right] \hat B^j
   \nonumber \\
   & & \qquad
       - {\sqrt{\overline h} \over N} \eta_{ijk} N^j
       \overline h{}^{k\ell}
       \hat E_\ell
       \equiv \hat B_i - \hat M_i^{\rm E},
   \nonumber \\
   & & \hat B^i = {\sqrt{\overline h} \over N} \overline h{}^{ij}
       \left( \breve B_j + \eta_{jk\ell} N^k \breve E{}^\ell
       \right) \equiv \breve B^i + \breve P^i_{\rm B},
   \nonumber \\
   & & \hat E_i = {N \over \sqrt{\overline h}}
       \left[ \left( 1 - {N^k N_k \over N^2} \right) \overline h_{ij}
       + {N_i N_j \over N^2} \right] \breve E{}^j
   \nonumber \\
   & & \qquad
       + {\sqrt{\overline h} \over N} \eta_{ijk} N^j
       \overline h{}^{k\ell}
       \breve B_\ell
       \equiv \breve E_i - \breve M_i^{\rm B},
   \label{EB-relation}
\eea
which Einstein called ``pretty complex relationship" \cite{Einstein-1916}.

Using these relations between the two definitions in Eq.\ (\ref{EB-relation}), two complete sets of Maxwell's equations can be written individually. In terms of $\hat E_i$ and $\hat B_i$, we have
\bea
   & & ( {\hat E}^i + {\hat P}_{\rm E}^i )_{,i}
       = {1 \over c} \sqrt{-g} J^0,
   \label{Maxwell-hat-1} \\
   & &
       ( {\hat E}^i + {\hat P}_{\rm E}^i )_{,0}
       - \eta^{ijk} \nabla_j ( {\hat B}_k
       - {\hat M}^{\rm E}_k )
       = - {1 \over c} \sqrt{-g} J^i,
   \label{Maxwell-hat-2} \\
   & & {\hat B}^i_{\;\;,i} = 0,
   \label{Maxwell-hat-3} \\
   & & {\hat B}^i_{\;\;,0}
       + \eta^{ijk} \nabla_j {\hat E}_k = 0,
   \label{Maxwell-hat-4}
\eea
where $\hat {\bf P}_{\rm E}$ and $\hat {\bf M}_{\rm E}$ are defined in Eq.\ (\ref{EB-relation}). The effective ${\bf P}$ and ${\bf M}$ appear only in the inhomogeneous part of Maxwell's equations, and in such a case the effect of gravity can be interpreted as the effective medium property.

In terms of $\breve {\bf B}$ and $\breve {\bf E}$, however, Eqs.\ (\ref{Maxwell-breve-1})-(\ref{Maxwell-hat-4}) give
\bea
   & & {\breve E}^i_{\;\;,i}
       = {1 \over c} \sqrt{-g} J^0,
   \label{Maxwell-breve-1} \\
   & &
       {\breve E}^i_{\;\;,0}
       - \eta^{ijk} \nabla_j {\breve B}_k
       = - {1 \over c} \sqrt{-g} J^i,
   \label{Maxwell-breve-2} \\
   & & ( {\breve B}^i + {\breve P}_{\rm B}^i )_{,i} = 0,
   \label{Maxwell-breve-3} \\
   & & ( {\breve B}^i + {\breve P}_{\rm B}^i )_{,0}
       + \eta^{ijk} \nabla_j ( {\breve E}_k
       - {\breve M}^{\rm B}_k ) = 0,
   \label{Maxwell-breve-4}
\eea
where $\breve {\bf P}_{\rm B}$ and $\breve {\bf M}_{\rm B}$ are defined in Eq.\ (\ref{EB-relation}). The effective ${\bf P}$ and ${\bf M}$ now appear in the homogeneous part which is opposite to the previous choice.

The above two non-covariant definitions are widely used in the literature of handling the effect of gravitational waves  \cite{Cooperstock-1968, Baroni-1985, Berlin-2022, Domcke-2022}, and interpreting the effect of gravity as a medium property \cite{Moller-1952, Skrotskii-1957, Balazs-1958, Plebanski-1960, deFelice-1971, Landau-Lifshitz-1971, Leonhardt-2006}. In these non-covariant definitions one cannot identify the four-vectors of corresponding observers who may measure such EM fields. Consequently, one cannot introduce the corresponding charge and current densities. We will elaborate our criticism on previous literature in Secs.\ \ref{sec:misconception} and \ref{sec:medium}.

Einstein in \cite{Einstein-1916} used these two separate coincidences with special relativity as the guide to suggest the covariant form of Maxwell's equations in (\ref{Maxwell-tensor-eqs}). However, he suggested non of these two as the definition of the EM fields \cite{Einstein-1916}.

%
%
%
\section{Generic observer}
                                         \label{sec:generic}

In the literature of gravitational wave detection using electromagnetic means, a special choice of the EM fields, {\it assuming} $F_{ab}$ (with two covariant indices) independent of the metric perturbations, is popular \cite{Cooperstock-1968, Baroni-1985, Berlin-2022, Domcke-2022}. In this way, the homogeneous part of Maxwell's equations becomes independent of gravity with apparent merit: the perturbed metric (including gravitational waves) appears only as the effective polarization and magnetization in ordinary Maxwell's equations in flat spacetime, similarly as in the axion case.

In the following, using a generic observer we will show that in the curved spacetime, such an {\it ad hoc} choice of the EM fields in Eqs.\ (\ref{Fab-hat}) or (\ref{Fab-breve}) is {\it not} possible for any observer. For the gravitation, it is inevitable to have effective polarizations and magnetizations appearing in {\it both} the homogeneous and inhomogeneous parts of Maxwell's equations. Gravitational wave detection should take this complication compared with the axion into account; for correct equations in the normal frame, see \cite{Hwang-Noh-2023-EM-definition}.

The fluid four-vector of a generic observer is introduced as \cite{Hwang-Noh-2022-Axion-EM}
\bea
   & & u_i \equiv \gamma V_i, \quad
       u_0 = \gamma \left( N_i V^i - N \right),
   \nonumber \\
   & &
       u^i = \gamma \left( V^i - {1 \over N} N^i \right) \equiv {\gamma \over N} \overline V^i, \quad
       u^0 = {1 \over N} \gamma,
   \label{u_a-ADM}
\eea
with the Lorentz factor
\bea
   \gamma \equiv - n_c u^c
       = N u^0
       = {1 \over \sqrt{1 - V^k V_k}},
   \label{Lorentz}
\eea
where indices of $V^i$ and $\overline V{}^i$ are raised and lowered by $\overline h_{ij}$ and its inverse; $V_i$ is the fluid velocity measured by the Eulerian observer with $n_a$ \cite{Smarr-York-1978, Wilson-Mathews-2003, Gourgoulhon-2012}, and $\overline V_i \equiv {u^i / u^0} = {d x^i / d x^0}$ is the coordinate velocity of the fluid measured by the coordinate observer \cite{Wilson-Mathews-2003, Baumgarte-Shapiro-2010}. For $V_i = 0$ we have the normal frame of an Eulerian observer, and for $\overline V_i \equiv 0$ we have the coordinate frame $\bar n_a$.

We introduce $\widetilde B_i \equiv \overline B_i$ with the index of $\overline B_i$ associated with $\overline h_{ij}$ as the metric. We have
\bea
   & & \widetilde B_i \equiv \overline B_i, \quad
       \widetilde B_0 = ( N^i - N V^i ) \overline B_i,
   \nonumber \\
   & &
       \widetilde B^i = \overline B^i
       - {N^i \over N} V^j \overline B_j, \quad
       \widetilde B^0 = {1 \over N} V^i \overline B_i,
\eea
and similarly for $\overline E_i$, $\overline D_i$, $\overline H_i$, and $\overline j_i$.

Equations (\ref{Maxwell-DH}) and (\ref{Maxwell-EB}), with $a = 0$ and $i$, respectively, give
\begin{widetext}
\bea
   & & {1 \over \sqrt{\overline h}} \left[ \sqrt{\overline h}
       \gamma \left( \overline D{}^i
       - V^i V^j \overline D_j
       - \overline \eta{}^{ijk} V_j \overline H_k
       \right) \right]_{,i}
       = \gamma \varrho + {1 \over c} V^i \overline j_i,
   \label{Maxwell-generic-1} \\
   & & {1 \over \sqrt{\overline h}} \left[ \sqrt{\overline h}
       \gamma \left( \overline D{}^i
       - V^i V^j \overline D_j
       - \overline \eta{}^{ijk} V_j \overline H_k
       \right) \right]_{,0}
       - \overline \eta{}^{ijk} \overline \nabla_j \Big\{
       \gamma \big[ ( N - N_\ell V^\ell ) \overline H_k
       + ( N^\ell - N V^\ell ) \overline H_\ell V_k \big]
   \nonumber \\
   & & \qquad
       - \gamma \overline \eta_{k\ell m} \left[
       ( N^\ell - N V^\ell ) \overline D{}^m
       - V^\ell N^m V^n \overline D_n \right] \Big\}
       = \gamma \varrho ( N^i - N V^i )
       - {1 \over c} ( N \overline j{}^i
       - N^i V^j \overline j_j ),
   \label{Maxwell-generic-2} \\
   & & {1 \over \sqrt{\overline h}} \left[ \sqrt{\overline h}
       \gamma \left( \overline B{}^i
       - V^i V^j \overline B_j
       + \overline \eta{}^{ijk} V_j \overline E_k
       \right) \right]_{,i}
       = 0,
   \label{Maxwell-generic-3} \\
   & & {1 \over \sqrt{\overline h}} \left[ \sqrt{\overline h}
       \gamma \left( \overline B{}^i
       - V^i V^j \overline B_j
       + \overline \eta{}^{ijk} V_j \overline E_k
       \right) \right]_{,0}
       + \overline \eta{}^{ijk} \overline \nabla_j \Big\{
       \gamma \big[ ( N - N_\ell V^\ell ) \overline E_k
       + ( N^\ell - N V^\ell ) \overline E_\ell V_k \big]
   \nonumber \\
   & & \qquad
       + \gamma \overline \eta_{k\ell m} \left[
       ( N^\ell - N V^\ell ) \overline B{}^m
       + V^\ell N^m V^n \overline B_n \right] \Big\}
       = 0.
   \label{Maxwell-generic-4}
\eea
\end{widetext}
The left-hand sides of Eqs.\ (\ref{Maxwell-generic-1})-(\ref{Maxwell-generic-4}) can be arranged so that the effect of gravity can be interpreted as the effective polarizations and magnetizations appearing in {\it both} the homogeneous and inhomogeneous parts as
\bea
   & & \hskip -.8cm
       {1 \over \sqrt{\overline h}} \left[ \sqrt{\overline h}
       ( \overline E{}^i + \overline P{}_{\rm E}^i ) \right]_{,i}
       = \gamma \varrho + {1 \over c} V^i \overline j_i,
   \label{Maxwell-g-1} \\
   & & \hskip -.8cm
       {1 \over \sqrt{\overline h}} \left[ \sqrt{\overline h}
       ( \overline E{}^i + \overline P{}_{\rm E}^i ) \right]_{,0}
       - \overline \eta^{ijk} \overline \nabla_j
       ( \overline B_k - \overline M{}^{\rm E}_k )
   \nonumber \\
   & & \qquad \hskip -.8cm
       = \gamma \varrho ( N^i - N V^i )
       - {1 \over c} ( N \overline j{}^i
       - N^i V^j \overline j_j ),
   \label{Maxwell-g-2} \\
   & & \hskip -.8cm
       {1 \over \sqrt{\overline h}} \left[ \sqrt{\overline h}
       ( \overline B{}^i + \overline P{}_{\rm B}^i ) \right]_{,i} = 0,
   \label{Maxwell-g-3} \\
   & & \hskip -.8cm
       {1 \over \sqrt{\overline h}} \left[ \sqrt{\overline h}
       ( \overline B{}^i + \overline P{}_{\rm B}^i ) \right]_{,0}
       + \overline \eta^{ijk} \overline \nabla_j
       ( \overline E_k - \overline M{}^{\rm B}_k ) = 0,
   \label{Maxwell-g-4}
\eea
where indices of $\overline P{}_{\rm E}^i$, etc., are associated with $\overline h_{ij}$. The effective $\overline {\bf P}$s and $\overline {\bf M}$s caused by the metric and the axion are
\bea
   & & \hskip -.8cm
       \overline P{}_{\rm E}^i
       \equiv \gamma \left( \overline D{}^i
       - V^i V^j \overline D_j
       - \overline \eta{}^{ijk} V_j \overline H_k \right)
       - \overline E{}^i,
   \nonumber \\
   & & \hskip -.8cm
       \overline M{}^{\rm E}_i
       \equiv - \gamma \left[ ( N - N_j V^j ) \overline H_i
       + ( N^j - N V^j ) \overline H_j V_i \right]
   \nonumber \\
   & & \qquad \hskip -.8cm
       + \gamma \overline \eta_{ijk} \left[
       ( N^j - N V^j ) \overline D{}^k
       - V^j N^k V^\ell \overline D_\ell \right]
       + \overline B_i,
   \nonumber \\
   & & \hskip -.8cm
       \overline P{}_{\rm B}^i
       \equiv \gamma \left( \overline B{}^i
       - V^i V^j \overline B_j
       + \overline \eta{}^{ijk} V_j \overline E_k \right)
       - \overline B{}^i,
   \nonumber \\
   & & \hskip -.8cm
       \overline M{}^{\rm B}_i
       \equiv - \gamma \left[ ( N - N_j V^j ) \overline E_i
       + ( N^j - N V^j ) \overline E_j V_i \right]
   \nonumber \\
   & & \qquad \hskip -.8cm
       - \gamma \overline \eta_{ijk} \left[
       ( N^j - N V^j ) \overline B{}^k
       + V^j N^k V^\ell \overline B_\ell \right]
       + \overline E_i.
   \label{PM-generic}
\eea
The effect due to axion is included in $\overline D_i$ and $\overline H_i$.

\subsection{The non-covariant definitions do not lead to measurable EM fields}

We will show that in the curved spacetime, the non-covariant definitions of the EM fields in Eqs.\ (\ref{Fab-hat}) or (\ref{Fab-breve}) are not possible for any observer. As a consequence, we will show that gravity inevitably causes modifications in both the homogeneous and inhomogeneous equations. From Eqs.\ (\ref{Fab-cov}) and (\ref{Hab-cov}), we have
\bea
   & & \hskip -.8cm
       F_{0i}
       = - \gamma ( N - N_j V^j ) \overline E_i
       - \gamma V_i ( N^j - N V^j ) \overline E_j
   \nonumber \\
   & & \qquad \hskip -.8cm
       + \gamma \overline \eta_{ijk} \left[ ( N V^j - N^j )
       \overline B{}^k
       - V^j N^k V^\ell \overline B_\ell \right],
   \nonumber \\
   & & \hskip -.8cm
       F_{ij}
       = \gamma ( V_i \overline E_j - V_j \overline E_i )
       + \gamma \overline \eta_{ijk} ( \overline B{}^k
       - V^k V^\ell \overline B_\ell ),
\eea
where the choice of observer's frame corresponds to choosing $V_i$. These conditions relate the six $F_{ab}$ to the six EM fields. We cannot achieve arbitrary relations between them using only three conditions available in $V_i$, which turns out to be the case for achieving $F_{ab}$ (or whatever combinations including $F^*_{ab}$) independent of the metric perturbations even to the linear order \cite{Hwang-Noh-2023-EM-definition}.

To the linear order in metric perturbations and $V_i$, we have
\bea
   & & F_{0i} = - N \overline E_i
       + \overline \eta_{ijk} ( N V^j - N^j )
       \overline B{}^k,
   \nonumber \\
   & &
       F_{ij}
       = V_i \overline E_j - V_j \overline E_i
       + \overline \eta_{ijk} \overline B{}^k.
\eea
Thus, even to the linear order in metric perturbations, one {\it cannot} remove the metric dependence of $F_{ab}$ by choosing $V_i$. This conclusion is independent of the gauge conditions. For example, for the normal frame with $V_i \equiv 0$, synchronous gauge condition with $N = 1$ and $N_i = 0$ still leaves dependence on $\sqrt{\overline h}$ and $\overline h{}^{ij}$ through $\overline \eta_{ijk}$ and $\overline B{}^i$.

Therefore, the non-covariant definitions of the EM fields introduced in \cite{Moller-1952, Landau-Lifshitz-1971, Berlin-2022, Domcke-2022}, na\"ively assigning the EM fields to components of the field strength tensor as in Eq.\ (\ref{Fab-hat}) cannot be regarded as any measurable EM fields. That is, as these variables cannot be measured as EM fields by any observer with their own four-vector, these are disqualified as the EM field variables. We can similarly show the non-covariant definition in Eq.\ (\ref{Fab-breve}) is not allowed.

\subsection{Using flat spacetime variables}

In terms of variables associated with $\delta_{ij}$, we introduce $\overline B_i \equiv B_i$ with the index of $B_i$ associated with $\delta_{ij}$, thus $\overline B{}^i = \overline h{}^{ij} B_j$, and similarly for $E_i$, $D_i$, $H_i$, and $j_i$. We also introduce $V_i \equiv v_i/c$ with the index of $v_i$ associated with $\delta_{ij}$, thus $V^i = \overline h{}^{ij} v_j/c$. Equations (\ref{Maxwell-g-1})-(\ref{PM-generic}) become
\bea
   & & ( E^i + P_{\rm E}^i )_{,i}
       = \sqrt{\overline h}
       \left( \gamma \varrho
       + {1 \over c} V^i j_i \right),
   \label{Maxwell-g-FNLE-1} \\
   & & ( E^i + P_{\rm E}^i )_{,0}
       - \eta^{ijk} \partial_j ( B_k - M^{\rm E}_k )
   \nonumber \\
   & & \qquad
       = \sqrt{\overline h}
       \left[ \gamma \varrho ( N^i - N V^i )
       - {1 \over c} ( N \overline j{}^i
       - N^i V^j j_j ) \right],
   \label{Maxwell-g-FNLE-2} \\
   & & ( B^i + P_{\rm B}^i )_{,i} = 0,
   \label{Maxwell-g-FNLE-3} \\
   & & ( B^i + P_{\rm B}^i )_{,0}
       + \eta^{ijk} \partial_j ( E_k - M^{\rm B}_k ) = 0,
   \label{Maxwell-g-FNLE-4}
\eea
where the effective $\overline {\bf P}$s and $\overline {\bf M}$s caused by the metric and the axion are
\bea
   & & P^i_{\rm E} = \sqrt{\overline h} \overline h{}^{ij}
       ( E_j + \overline P{}^{\rm E}_j ) - E^i, \quad
       M^{\rm E}_i = \overline M{}^{\rm E}_i,
   \nonumber \\
   & & P^i_{\rm B} = \sqrt{\overline h} \overline h{}^{ij}
       ( B_j + \overline P{}^{\rm B}_j ) - B^i, \quad
       M^{\rm B}_i = \overline M{}^{\rm B}_i.
   \label{PM-generic-FNLE}
\eea
For $v_i \equiv 0$ we recover Eqs.\ (\ref{Maxwell-normal-1})-(\ref{PM-normal}) in the normal frame, and for $V^i \equiv N^i/N$ we recover Eqs.\ (\ref{Maxwell-coordinate-1})-(\ref{Maxwell-coordinate-4}) in the coordinate frame.

\subsection{Linear order metric perturbations}

To the linear order, using Eq.\ (\ref{ADM-linear}), we have
\bea
   & & \hskip -.8cm
       F_{0i}
       = - \left( 1 + {1 \over 2} h^0_0 \right) E_i
       - \eta_{ijk} \left( h^j_0 - {v^j \over c} \right) B^k,
   \nonumber \\
   & & \hskip -.8cm
       F_{ij}
       = {v_i \over c} E_j - {v_j \over c} E_i
       + \eta_{ijk} \left[ \left( 1
       + {1 \over 2} h^\ell_\ell \right) B^k
       - h^{k\ell} B_\ell \right],
   \label{Fab-g-linear}
\eea
and
\bea
   & & ( E^i + P_{\rm E}^i )_{,i}
       = \varrho \left( 1 + {1 \over 2} h^i_i \right)
       + {v^i \over c^2} j_i,
   \label{Maxwell-g-linear-1} \\
   & &
       ( E^i + P_{\rm E}^i )_{,0}
       - \eta^{ijk} \nabla_j
       ( B_k- M^{\rm E}_k )
       =
       \left( h^i_0 - {v^i \over c} \right) \varrho
   \nonumber \\
   & & \qquad
       - {1 \over c} \left( 1 + {1 \over 2} h^j_j
       + {1 \over 2} h^0_0 \right) j^i
       + {1 \over c} h^{ij} j_j,
   \label{Maxwell-g-linear-2} \\
   & & ( B^i + P_{\rm B}^i )_{,i} = 0,
   \label{Maxwell-g-linear-3} \\
   & & ( B^i + P_{\rm B}^i )_{,0}
       + \eta^{ijk} \nabla_j ( E_k - M^{\rm B}_k )
       = 0,
   \label{Maxwell-g-linear-4}
\eea
with
\bea
   & & P_{\rm E}^i
       \equiv D^i - E^i + {1 \over 2} h^j_j D^i - h^{ij} D_j
       - \eta^{ijk} {v_j \over c} H_k,
   \nonumber \\
   & & M_{\rm E}^i
       \equiv B^i - H^i - {1 \over 2} h^0_0 H^i
       + \eta^{ijk} \left( h_{0j} - {v_j \over c} \right) D_k,
   \nonumber \\
   & & P_{\rm B}^i
       \equiv {1 \over 2} h^j_j B^i - h^{ij} B_j
       + \eta^{ijk} {v_j \over c} E_k,
   \nonumber \\
   & & M_{\rm B}^i
       \equiv - {1 \over 2} h^0_0 E^i
       - \eta^{ijk} \left( h_{0j} - {v_j \over c} \right) B_k.
   \label{PM-g-linear}
\eea
From Eq.\ (\ref{Fab-g-linear}) we can see that it is not possible to remove dependence on the metric perturbation ($h_{ab}$) by choosing a frame ($v_i$) \cite{Hwang-Noh-2023-EM-definition}. For $v_i = 0$ and $v_i/c = h_{0i}$ we recover the equations in the normal frame and the coordinate frame, respectively \cite{Hwang-Noh-2023-EM-definition}.

%
%
%
\section{Misconceptions in the literature}
                                   \label{sec:misconception}

There are two {\it wrong} arguments \cite{Berlin-2022, Domcke-2022} used to support the absence of metric in the definition of EM fields using $F_{ab}$ with two covariant indices; equivalently, the absence of metric in the homogeneous Maxwell's equations.

Authors of \cite{Berlin-2022} argued for metric independence of the homogeneous Maxwell's equation using the topological nature of the equation in terms of form algebra. On this regards, however, the inhomogeneous equation is also topological, with no relation to metric. According to \cite{MTW-1973}, ``Remarkably, neither equation makes any reference whatsoever to {\it metric}. [T]he concepts of form and exterior derivative are metric-free. Metric made an appearance only in one place, in the concept of duality (``perpendicularity") that carried attention from ${\bf F}$ to the dual structure ${\bf F}^*$." Thus, the authors of \cite{Berlin-2022} confused the topological nature of the homogeneous Maxwell's equation as the absence of metric in the EM field decomposition. The metric appears when we related the two parts, and also in the decomposition of field strength tensor into the EM fields.

Authors of \cite{Domcke-2022} similarly argued for metric independence of $F_{ab}$ using the absence of the metric in the relation between $F_{ab}$ and the four-potential $A_a$, i.e., \bea
   F_{ab} \equiv A_{b,a} - A_{a,b}.
   \label{four-potential}
\eea
This, however, does not imply the absence of metric between $F_{ab}$ and the EM fields. In the normal-frame, Eq.\ (\ref{Fab-normal}) can be inverted to gives
\bea
   E_i = - {1 \over N} F_{0i} - {N^j \over N} F_{ij}, \quad
       B_i = {1 \over 2 \sqrt{\overline h}} \overline h_{ij}
       \eta^{jk\ell} F_{k\ell}.
\eea
Using Eq.\ (\ref{four-potential}), we have
\bea
   & & E_i = {1 \over N} \left[ \partial_i A_0
       - A_{i,0}
       - ( \partial_i A_j - \partial_j A_i ) N^j \right],
   \nonumber \\
   & & B_i
       = {1 \over \sqrt{\overline h}}
       \overline h_{ij} \eta^{jk\ell} \partial_k A_\ell.
   \label{EM-potenial-normal}
\eea
Thus, metric is heavily involved in the relation between the EM fields and four-potential. Only in Minkowski spacetime, we recover the well-known relations ${\bf E} = \nabla A_0 - {\bf A}_{,0}$ and ${\bf B} = \nabla \times {\bf A}$.

Thus, the authors of \cite{Domcke-2022} confused the absence of metric in its relation to the four-potential as the absence of the metric in the EM field decomposition of $F_{ab}$. The second one (homogeneous part) in Eq.\ (\ref{Maxwell-tensor-eqs}) is naturally satisfied by using $A_a$; i.e., written in terms of the four-potential, the homogeneous part is identically satisfied without metric influence. However, in terms of the EM fields, the homogeneous part is still coupled with the metric because the EM fields are non-trivially related to the four-potential for any choice of the frame four-vector; see Eq.\ (\ref{EM-potenial-normal}) for the relation to the EM fields in the normal frame. As we show below, in curved spacetime the EM variables defined in the simple relations in Eq.\ (\ref{EM-potenial}) are nothing but $\hat E_i$ and $\hat B_i$ which do not correspond to measurable EM fields.

Authors of \cite{Baroni-1985} derived perturbed Maxwell's equations via variation of the four-potential $A_a$. As we mentioned, this corresponds to {\it assuming} $F_{ab}$ with two covariant indices independent of the metric which is fine if one uses $A_a$. If one associate the potential to the EM fields na\^ively using the special relativistic relations, the Maxwell's equations derived in this way correspond to the ones using $\hat E_a$ and $\hat B_a$, i.e.,
\bea
   \hat E_i = \partial_i A_0 - \partial_0 A_i, \quad
       \hat B_i = \eta_{ijk} \partial^j A^k.
   \label{EM-potenial}
\eea
Using Eq.\ (\ref{EM-potenial-normal}), we can show that the relation between ($E_i$, $B_i$) and ($\hat E{}_i$, $\hat B{}_i$) is the same as in Eq.\ (\ref{EB-relation-normal-2}). We have expounded in this work that $\hat E{}_i$ and $\hat B{}_i$ do not correspond to EM fields measured by any observer. Thus, the EM variables used in Maxwell's equations derived in such a way should be translated to the proper EM fields based on the observer's four-vector, for example, using Eq.\ (\ref{EB-relation-normal-2}) or (\ref{EM-potenial-normal}) for the Eulerian observer.

The presence of metric in the decomposition of $F_{ab}$ and $F^*_{ab}$ into the EM fields is partly related to the presence of metric in the four-vector in gravitational environments. The generic four-vector in Eq.\ (\ref{u_a-ADM}) shows that for any choice of $V_i$, it is not easy to have the special relativistic form of the normal four-vector $n_a = (-1, 0, 0, 0)$. In the normal frame, by imposing synchronous gauge condition with $N \equiv 1$ and $N_i \equiv 0$, we can achieve this form with $n^a = (1, 0, 0, 0)$. But, even in such a case, Eq.\ (\ref{Fab-normal}) shows that $F_{ij}$ still depends on the metric through $\overline h{}^{k\ell}$ and $\overline h$.

The presence of metric in the decomposition of $F_{ab}$ into EM fields directly leads to the presence of metric in the homogeneous part of Maxwell's equations. Therefore, proposals for gravitational wave detection using the axion haloscope in \cite{Berlin-2022, Domcke-2022} are based on {\it wrong} Maxwell's equations. Although the effect of axion can be treated as an effective medium, the effect of gravity is more complicated, as explained in this work. Proper study should use Maxwell's equations in the normal frame in Eqs.\ (\ref{Maxwell-normal-1-1})-(\ref{Maxwell-normal-1-4}) or Eqs.\ (\ref{Maxwell-normal-1})-(\ref{PM-normal}).

%
%
%
\section{Errors in medium interpretation}
                                   \label{sec:medium}

It is well known that the effect of axion coupling to electrodynamics in flat spacetime can be interpreted as an effective medium property in electrodynamics \cite{Sikivie-1983, Wilczek-1987, Sikivie-2021}. There are numerous works interpreting the effect of gravity similarly as an effective medium property. To our knowledge, however, {\it all} the previous literature on the subject are, in fact, based on the two non-covariant identifications of the EM fields \cite{Moller-1952, Skrotskii-1957, Balazs-1958, Plebanski-1960, deFelice-1971, Landau-Lifshitz-1971, Leonhardt-2006}, and thus are in error.

Most of the literature concerning medium interpretation of gravity follow Plebanski \cite{Plebanski-1960}. Plebanski {\it defined} the EM fields in vacuum as
\bea
   & & F_{0i} \equiv -  E_i, \quad
       F_{ij} \equiv \eta_{ijk} B^k,
   \label{Fab-medium-0} \\
   & & \sqrt{-g} F^{0i} \equiv D^i, \quad
       \sqrt{-g} F^{ij} \equiv \eta^{ijk} H_k,
   \label{Fab-medium-1}
\eea
where indices of $E_i$, $B_i$, $D_i$, and $H_i$ are associated with $\delta_{ij}$ and its inverse. Using Eq.\ (\ref{eta-definition}), Eq.\ (\ref{Fab-medium-1}) is the same as
\bea
   F^*_{0i} \equiv - H_i, \quad
       F^*_{ij} \equiv - \eta_{ijk} D^k.
   \label{Fab-medium-2}
\eea
Using Eqs.\ (\ref{Fab-medium-0}) and (\ref{Fab-medium-1}), Eq.\ (\ref{Maxwell-tensor-eqs}), ignoring the axion coupling, gives
\bea
   & & D^i_{\;\;,i} = {1 \over c} \sqrt{-g} J^0,
   \label{Maxwell-P-1} \\
   & & D^i_{\;\;,0} - \eta^{ijk} \nabla_j \hat H_k
       = - {1 \over c} \sqrt{-g} J^i,
   \label{Maxwell-P-2} \\
   & & B^i_{\;\;,i} = 0,
   \label{Maxwell-P-3} \\
   & & B^i_{\;\;,0} + \eta^{ijk} \nabla_j E_k = 0.
   \label{Maxwell-P-4}
\eea
These are ordinary Maxwell's equations in flat spacetime which Minkowski has derived in \cite{Minkowski-1907}; notice that we kept the four-current instead of the charge and current densities, see below.

In order to derive the effective constitutive relations, Plebanski used
\bea
   \sqrt{-g} F^{ab} = \sqrt{-g} g^{ac} g^{bd} F_{cd}.
   \label{H-F}
\eea
Using Eq.\ (\ref{Fab-medium-1}), we can show
\bea
   {D}^i
       = \varepsilon^{ij} E_j
       + \gamma^{ij} {H}_j, \quad
       B^i = \mu^{ij} {H}_j
       - \gamma^{ij} E_j,
\eea
with
\bea
   & & \varepsilon^{ij} = \mu^{ij}
       = {N \sqrt{\overline h} \over N^2 - N^k N_k}
       \left( \overline h^{ij} - {N^i N^j \over N^2} \right)
       = {\sqrt{-g} \over - g_{00}} g^{ij},
   \nonumber \\
   & &
       \gamma^{ij} = {\eta^{ijk} N_k \over
       N^2 - N^\ell N_\ell}
       = \eta^{ijk} {g_{0k} \over - g_{00}},
   \label{constitutive}
\eea
where $\varepsilon^{ij}$ and $\mu^{ij}$ are the permittivity and permeability tensors, respectively, and $\gamma^{ij}$ is the electromagnetic mixing term; indices of $\varepsilon_{ij}$, etc., are associated with $\delta_{ij}$ and its inverse.

Notice that definitions in Eq.\ (\ref{Fab-medium-0}) and Eq.\ (\ref{Fab-medium-2}) exactly correspond to the non-covariant definitions in Eqs.\ (\ref{Fab-hat}) and (\ref{Fab-breve}), respectively. Thus, ($E_i$, $B_i$) and ($D_i$, $H_i$) are nothing but our non-covariant definitions ($\hat E_i$, $\hat B_i$) and ($\breve D_i$, $\breve H_i$), respectively. As explained in previous sections, in these non-covariant ways of defining the EM fields, one {\it cannot} identify the four-vector allowing these definitions and consequently cannot introduce the accompanying charge and current densities, even to the linear order in gravity \cite{Hwang-Noh-2023-EM-definition}; this is why we kept the four-current in Eqs.\ (\ref{Maxwell-P-1}) and (\ref{Maxwell-P-2}).

The non-covariant definitions of EM fields introduced in \cite{Moller-1952, Landau-Lifshitz-1971} differ from the ones in Plebanski by using the three-space covariant metric $\overline \gamma {}_{ij}$, defined as $\overline \gamma{}^{ij} \equiv g^{ij}$, instead of $\delta_{ij}$ used in \cite{Plebanski-1960}; see Sec.\ 115 in \cite{Moller-1952} and Sec.\ 90 in \cite{Landau-Lifshitz-1971}. In this way, we can show that the constitutive relations (thus the medium property) differ from Plebanski's \cite{Hwang-Noh-EM-medium-2023}. Using the non-covariant definitions of EM fields in gravity systems, the {\it dependence} of medium property on the spatial decomposition method and the coordinate used are also noticed in \cite{Gibbons-Werner-2019}. We will analyse the errors in previous literature concerning the medium interpretation in detail in a subsequent work \cite{Hwang-Noh-EM-medium-2023}.

In Sec.\ \ref{sec:generic} we showed that the above non-covariant definitions do not have corresponding observer's four-vectors, thus these are not measurable EM fields by any observer. We also proved that any choice of the observer's four-vector leads to the presence of metric in the relation between the field strength tensor (in any form) and the EM fields. As a direct consequence, gravity causes modification of Maxwell's equations in both the homogeneous and inhomogeneous parts. The presence of metric in the homogeneous Maxwell's equations cannot be interpreted as the medium property in ordinary sense. In this way, the medium interpretation of gravity in previous literature is in error.

%
%
%
\section{Discussion}
                                   \label{sec:Discussion}

Our main results are Maxwell's equations in a general curved spacetime presented in forms valid in Minkowski background with effects of gravity appearing as effective ${\bf P}$s and ${\bf M}$s. The ADM metric variables are expressed in terms of the metric perturbation variables associated with $\delta_{ij}$ using the FNLE formulation, see Eq.\ (\ref{FNLE-metric}). We consider four different definitions of the EM fields, two based on the covariant decomposition, and two others using non-covariant identifications directly matching the EM fields with tensor components of $F_{ab}$ and $F^*_{ab}$. We also presented the case for the generic observer.

The two non-covariant definitions of EM fields lead to trouble as these fail to identify the observer's four-vectors, who may measure such fields, thus failing to identify the corresponding external charge and current densities, which also depend on the same four-vector (observer). This trouble was noticed even in the weak gravity case \cite{Hwang-Noh-2023-EM-definition}. Here, we prove that such definitions are {\it not} possible for the EM fields using the analysis for a generic observer. As a consequence, to any observer the gravity causes the homogeneous part of Maxwell's equations modified in addition to the inhomogeneous part. We propose the normal frame as the one to use as it corresponds to the Eulerian observer. Relations to the other definitions of the EM fields, including the non-covariant ones, are derived in this work.

In Laboratory situations, like in the Earth's gravitational field, in measuring passing gravitational waves using electromagnetic means, in inverse designing medium property for desired optical path using curved geometry, and in given weak gravitational fields, Maxwell's equations are enough with the gravity encoded in a given spacetime metric. The previous literature, however, rely on defining the EM fields in non-covariant manner. Here we showed that the EM fields defined in non-covariant manner are {\it not} the EM fields measured by any observer. Consequently, gravity causes modifications in both homogeneous and inhomogeneous parts, and the gravity-medium analogy is not valid. In the presence of gravity, for optical properties we should analyze Maxwell's equations directly. In the normal-frame, these are Eqs.\ (\ref{Maxwell-normal-1-1})-(\ref{Maxwell-normal-1-4}) or Eqs.\ (\ref{Maxwell-normal-1})-(\ref{PM-normal}) in general curved spacetime.

In realistic astrophysical situations, Maxwell's equations should be dynamically combined with Einstein's equation with accompanying fluids and other fields. The complete sets of Einstein's equation with a fluid and a scalar field (with additional helical coupling with the EM fields) in the covariant, the ADM, and the FNLE formulations are presented in \cite{Hwang-Noh-2022-Axion-EM}. Maxwell's equations in the general curved spacetime derived in different forms in this work are applicable in all these situations. The cosmological FNLE formulation of Maxwell's equations presented in \cite{Hwang-Noh-2022-Axion-EM} considered the normal frame and ignored the tensor-type perturbations. Cosmological extension of the present work using the FNLE formulation is trivial.

%
%
%
\section*{Acknowledgments}

We thank Anton Sokolov for useful discussion and Professor Gary Gibbons for clarifying comments. H.N.\ was supported by the National Research Foundation (NRF) of Korea funded by the Korean Government (No.\ 2018R1A2B6002466 and No.\ 2021R1F1A1045515). J.H.\ was supported by IBS under the project code, IBS-R018-D1, and by the NRF of Korea funded by the Korean Government (No.\ NRF-2019R1A2C1003031).

\appendix

%
%
%
\section{FNLE formulation}
                                        \label{sec:FNLE}

The FNLE formulation is designed to handle cosmological perturbations to fully nonlinear order \cite{Hwang-Noh-2013, Gong-2017}. The final equations are exact and the metric variables are directly visible in the equations. The essential point is that we can derive exact inverse metric for the most general nonlinear perturbations of the Friedmann background without imposing the gauge condition. In Minkowski background, simply setting the scale factor to unity in \cite{Hwang-Noh-2013, Gong-2017}, the FNLE metric convention is
\bea
   & &
       d s^2 = - ( 1 + 2 \alpha ) (d x^0)^2
       - 2 B_i d x^0 d x^i
       + \big[ ( 1 + 2 \varphi ) \delta_{ij}
   \nonumber \\
   & & \qquad
       + 2 \gamma_{,ij} + C_{i,j} + C_{j,i}
       + 2 C_{ij} \big] d x^i d x^j,
\eea
where $x^0 = ct$ and indices of $B_i$, $C_i$, and $C_{ij}$ are raised and lowered by $\delta_{ij}$ and its inverse; $C_i$ is transverse ($C^i_{\;\;,i} \equiv 0$), and $C_{ij}$ is symmetric, transverse, and trace-free ($C^j_{i,j} \equiv 0 \equiv C^i_i$). The perturbation variables may have arbitrary amplitudes only subject to the gravitational field equation. Without losing generality and mathematical convenience we can impose $\gamma \equiv 0 \equiv C_i$ as the spatial gauge condition; this is the unique spatial gauge condition which removes the spatial gauge degrees of freedom completely, thus the remaining variables (together with similar temporal gauge condition) are gauge invariant to fully nonlinear order in perturbations \cite{Hwang-Noh-2013, Noh-Hwang-2004}; in this gauge condition we set $B_i = \chi_i$.

The ADM metric quantities are derived in terms of the FNLE metric for the above general metric perturbations in \cite{Gong-2017}. Imposing $\gamma \equiv 0 \equiv C_i$ as the spatial gauge conditions, we have
\bea
   & & N = \sqrt{ 1 + 2 \alpha
       + {\delta^{ij} + H^{ij} \over a (1 + 2 \widehat \varphi)}
       \chi_i \chi_j },
   \nonumber \\
   & &
       N_i = - \chi_i, \quad
       N^i = - {\delta^{ij} + H^{ij} \over 1 + 2 \widehat \varphi} \chi_j,
   \nonumber \\
   & &
       \overline h_{ij} = ( 1 + 2 \varphi ) \delta_{ij} + 2 C_{ij}, \quad
       \overline h^{ij} = {\delta^{ij} + H^{ij} \over 1 + 2 \widehat \varphi},
   \nonumber \\
   & &
       \overline h \equiv {\rm det}(\overline h_{ij})
       = [ ( 1 + 2 \varphi )^2 - 2 C^{ij} C_{ij} ]
       ( 1 + 2 \widehat \varphi ),
   \nonumber \\
   & &
       H^{ij} \equiv - 2 { (1 + 2 \varphi) C^{ij}
       - 2 C^{ik} C^j_k \over (1 + 2 \varphi)^2
       - 2 C^{\ell m} C_{\ell m}},
   \nonumber \\
   & &
       \widehat \varphi \equiv \varphi
       + {2 \over 3} { 2 C_{ij} C^i_k C^{jk} \over
       (1 + 2 \varphi)^2 - 2 C^{\ell m} C_{\ell m}}.
   \label{FNLE-metric}
\eea

For the general case without imposing the spatial gauge condition, see Sec.\ 2 of \cite{Gong-2017}. Notice that we have not imposed the temporal gauge (hypersurface or slicing) condition; for several gauge conditions valid to fully nonlinear order, see \cite{Noh-Hwang-2004, Hwang-Noh-2013, Gong-2017}. With these ADM metric quantities determined in terms of the metric variables associated with $\delta_{ij}$, Maxwell's equations in this work can be expressed as the ones in {\it flat} spacetime with the effect of gravity (metric) appearing as effective ${\bf P}$s and ${\bf M}$s.

%
%
%
\section{Linear perturbation}
                                        \label{sec:linear}

To the linear order perturbations in the metric, using the notation in \cite{Hwang-Noh-2023-EM-definition}, i.e., $g_{ab} \equiv \eta_{ab} + h_{ab}$ with indices of $h_{ab}$ associated with $\eta_{ab}$ as the metric, we have
\bea
   h_{00} = - 2 \alpha, \quad
       h_{0i} = - \chi_i, \quad
       h_{ij} = 2 \varphi \delta_{ij} + 2 C_{ij},
\eea
thus
\bea
   & & \hskip -.5cm
       N = 1 - {1 \over 2} h_{00}, \quad
       N_i = h_{0i}, \quad
       N^i = h^i_0,
   \nonumber \\
   & & \hskip -.5cm
       \overline h_{ij} = \delta_{ij} + h_{ij}, \quad
       \overline h{}^{ij} = \delta^{ij} - h^{ij}, \quad
       \overline h = 1 + h^i_i.
   \label{ADM-linear}
\eea
These relations are valid even in the presence of $\gamma$ and $C_i$, thus for general linear metric perturbations without any condition. Using these relations between the FNLE metric and the linear perturbations, all the Maxwell's equations in this work reduce to the ones in \cite{Hwang-Noh-2023-EM-definition} which are valid to linear order in metric perturbations.

%
%


\end{document}